# Geochemistry of silicate-rich rocks can curtail spreading of carbon dioxide in subsurface aquifers

S. S. S. Cardoso* and J. T. H. Andres

*Department of Chemical Engineering & Biotechnology, University of Cambridge,*

*Cambridge CB2 3RA, UK*

*Electronic mail: sssc1@cam.ac.uk

**Abstract**

Pools of carbon dioxide are found in natural geological accumulations and in engineered storage in saline aquifers. It has been thought that once this $CO_2$ dissolves in the formation water, making it denser, convection streams will transport it efficiently to depth, but this may not be so. Here, we assess theoretically and experimentally the impact of natural chemical reactions between the dissolved $CO_2$ and the rock formation on the convection streams in the subsurface. We show that, while in carbonate rocks the streaming of dissolved carbon dioxide persists, the chemical interactions in silicate-rich rocks may curb this transport drastically and even inhibit it altogether. These results challenge our view of carbon sequestration and dissolution rates in the subsurface, suggesting that pooled carbon dioxide may remain in the shallower regions of the formation for hundreds to thousands of years. The deeper regions of the reservoir can remain virtually carbon free.

A large natural flux of carbon occurs in the Earth's subsurface, having a global magnitude of the order of 0.8 GtC per year[1]. Contributing to this flux, and to the carbon cycle, is the $CO_2$ transported from natural pools in a range of geological settings, particularly sedimentary basins, intra-plate volcanic regions and faulted areas[2]. More recently, the geological storage of anthropogenic carbon dioxide in subsurface saline aquifers has been explored as a technique to reduce emissions into the atmosphere[2-8]. An important challenge in the development of this technology is to predict the behaviour and eventual fate of the stored carbon dioxide. When relatively pure, whether as a gas, liquid or in supercritical form, carbon dioxide is in general less dense than formation water and will therefore rise in the subsurface as a result of buoyancy forces. However, once it dissolves in the formation water, the mixture becomes denser. Thus, a dense boundary layer of carbon-rich water grows by diffusion. This layer may later become gravitationally unstable, breaking up into streaming currents (termed fingers in fluid mechanics) that transport the $CO_2$-rich water downward to increasing depths, while deeper formation water rises[9-11]. Convection is a much more efficient mode of transport than diffusion, inducing downward fluxes of water with dissolved carbon dioxide of magnitude $u \sim k\,\Delta\rho_0\,g/\mu$, which can be as large as 1 m³ per m² in a year, in a saturated rock with permeability $k = 10^{-13}$ m², for example; here, $\Delta\rho_0$ is the density contrast between $CO_2$-saturated saline formation water (brine) and pure brine, $g$ is the acceleration of gravity and $\mu$ is the viscosity of the brine. This mechanism of carbon transport has attracted much scientific attention in the context of geological storage, because it enhances the dissolution of the pooled carbon dioxide and thereby increases the long-term stability and safety of storage. Experimental[12-16], theoretical[10,12,17-19] and numerical[10,20-23] studies have calculated the time for onset of convection, the evolution of the thickness of the $CO_2$-rich layer and the dissolution rate of the $CO_2$ pool.

Much less researched is the role of natural geochemical reactions between $CO_2$-rich, acidic brine and the host rock on the dissolution and spreading of carbon dioxide in the subsurface. For example, the dissolution reaction between the acidic brine and a rock formation rich in calcite,

$$2CaCO_3(s) + CO_2(aq) + H_2O(l) \rightarrow CO_3^{2-}(aq) + 2Ca^{2+}(aq) + 2HCO_3^{-}(aq), \qquad (1)$$

forms a solution denser than the acidic brine[1,24]. This enhances convection and the transport of $CO_2$-rich brine to great depth in the medium. Conversely, a precipitation reaction such as that between the acidic brine and a rock formation rich in calcium feldspar,

$$CaAl_2Si_2O_8(s) + CO_2(aq) + 2H_2O(l) \rightarrow CaCO_3(s) + Al_2Si_2O_5(OH)_4(s), \qquad (2)$$

promotes the deposition of solid calcite and kaolinite, removing carbon dioxide from the liquid phase[1,24]. Such a reaction will attenuate convective motion. The impact of geochemical reactions such as these depends on the relative timescales for fluid motion, diffusion and reaction. Earlier work[25] suggested timescales of hundreds of years for convection, millions of years for diffusion; and weeks to centuries for slow and fast reactions, respectively (for a reservoir of depth 100 m and permeability $10^{-13}$ m²). In the limit of very slow reactions, the geochemistry has little impact on fluid motion. However, intermediate and fast reactions may have a dramatic effect on the transport of the stored carbon dioxide. Previous numerical work[25-28] has indeed noted the possibility of a delay in the onset of convection caused by reaction, but has been unable to address the mechanisms by which reaction may suppress convection. It has moreover been assumed that, once initiated, convection would continue indefinitely in an unbounded domain.

In this article, we consider a chemical reaction between the dissolved $CO_2$ and a rock mineral $M_1$ (*e.g.*, calcium feldspar) that removes carbon from the formation brine[24], reducing the mixture density, and thereby curbing the complex buoyant fingering. We show how such a reaction can change profoundly the hydrodynamic behaviour of a diffusive layer of carbon dioxide and ultimately inhibit convection. To do so, we conduct a stability analysis of a transient diffusive layer undergoing a first-order, irreversible chemical reaction and compare the theoretical results with those of new laboratory experiments. We show here that the stability of a boundary layer depends crucially on how reaction constrains the wavelength of a random perturbation and on its effect on the diffusive concentration profile of the $CO_2$.

**Results**

**Dimensionless parameters from fluid dynamics and chemistry.** We consider an infinite, two-dimensional homogeneous porous medium where pure solute A ($CO_2$) overlies an aqueous layer (see Figure 1). The solute diffuses into the underlying aqueous layer forming a growing boundary layer. A first-order reaction $A(aq) + M_1(s) \rightarrow M_2(s)$ occurs between the dissolved solute A and a chemical species $M_1$ in the porous solid, with reaction rate $k_r a [T^{-1}]$, where the solid-based kinetic rate constant is $k_r [\text{mol} \cdot L^{-2} T^{-1}]$ and the reactive surface area of the solid is $a [L^2 \cdot \text{mol}^{-1}]$. The system is described by the Darcy-Boussinesq equations (see Methods). Scaling of these equations with appropriate boundary conditions shows that, for a reservoir of large depth compared to the solutal layer thickness, the only parameter determining flow and transport is a ratio of the Damköhler number and the square of the solutal Rayleigh-Darcy number, $\text{Da}/\text{Ra}^2 = k_r a D \phi / (k \Delta \rho_0 g / \mu)^2$. Here $\Delta \rho_0 [ML^{-3}]$ is the maximum density contrast between pure and solute-saturated fluid, $D[L^2 T^{-1}]$ is the diffusivity of the solute in the aqueous layer, $\varphi$ is the porosity of the rock matrix, and $g[LT^{-2}]$ is the acceleration of gravity. The permeability of the porous medium $k[L^2]$ and the fluid viscosity $\mu [ML^{-1}T^{-1}]$ are constant. The parameter $\text{Da}/\text{Ra}^2$ is a measure of the relative magnitude between the timescale for onset of convection and the timescale for reaction.

Thus, a small value of $Da/Ra^2$ reflects a slow reaction with little effect on the development of convection. Contrariwise, a large value of $Da/Ra^2$ implies a fast reaction with impact on the intensity of motion.

**Theoretical.** The linear stability of the system above was analysed with respect to transverse perturbations of wavenumber $k$ in the concentration and velocity fields. Figure 2 shows non-dimensionally the marginal stability curves (growth rate $\sigma = 0$) for different values of $Da/Ra^2$. Inside each U-shaped or closed curve the diffusive layer is unstable to transverse disturbances ($\sigma > 0$), while in the outer region it is stable ($\sigma < 0$). The diffusive layer is stable at early times but at larger times there is a range of unstable modes, between a small and a large wavenumber cut-off. For the inert system ($Da/Ra^2 = 0$), the time for onset of convection is almost constant at intermediate values of $k$, owing to the stabilizing effect of vertical diffusion. Vertical diffusion has a relative maximum influence at small wavenumbers when transverse diffusion is negligible, leading to an increase in the time for onset of convection on the left side of Figure 2. At high wavenumbers, transverse diffusion is dominant and stabilizes the system. Reaction has a stabilizing effect for all wavenumbers, thereby increasing the time for onset of convection, and concomitantly causing an increase in the low wavenumber cut-off and a decrease in the high wavenumber cut-off, with a consequent narrowing of the window of instability. Crucially, as the destabilizing concentration gradient weakens with time, reaction can overcome its effect and stabilize the system after a finite period of convection, as reflected by the closed marginal curves. This finite period of growth of instability decreases as $Da/Ra^2$ increases until it disappears at the critical value $Da/Ra^2 = 2.69 \times 10^{-3}$, above which the system becomes unconditionally stable for all wavenumbers at all times.

The non-dimensional maximum growth rates in Figure 3 exhibit a rapid increase at small times followed by a slower decay after the maximum. Reaction has two stabilizing effects: it reduces the concentration perturbation by consuming solute and it decreases the magnitude of the destabilizing base-concentration gradient. These mechanisms have the largest joint effect on the most unstable wavenumber, causing a pronounced decrease in the maximum growth rate as the speed of reaction rises. For the critical value of $Da/Ra^2 = 2.69 \times 10^{-3}$, all wavenumbers are stable. The most unstable wavenumber is independent of $\mathrm{Da}/\mathrm{Ra}^2$ and decreases gradually with time tending to an asymptotic value of $k = 0.05$. This temporal shift to lower wavenumbers is caused by the weaker effect of vertical diffusion compared to transverse diffusion as time progresses. Since reaction delays the onset of convection, the temporal shift in $k$ causes a decrease in the wavenumber at onset of convection as reaction speed increases.

**Laboratory Experiments.** To validate our theoretical predictions, new laboratory experiments using an acid-base reaction in a Hele-Shaw cell were developed. Although the fingering of acid-base reactions has been studied before for immiscible[29,30] and miscible[31-33] layers, these systems cannot mimic the $CO_2$ problem described above, because either convection developed in both layers or reaction was limited to the surface between the two layers. To mimic the growing boundary layer of carbon dioxide, we devised a new reaction system in which a dense solute diffuses from the upper layer into an immiscible lower layer where it undergoes a first-order reaction. Experiments were performed in a Hele-Shaw cell with a layer of phenolphthalein dissolved in 4-methyl-2-pentanone (commonly known as methyl isobutyl ketone or MIBK) overlying an aqueous solution of sodium hydroxide. Colourless phenolphthalein ($H_2P$) is instantaneously ionized in the presence of a strong base to form the dark pink anion $P^{2-}$; the anion then more slowly reacts with the strong base (pH>11) to form colourless carbinol $POH^{3-}$, according to[34],

$$\begin{aligned} H_2P(l) + 2OH^-(aq) &\rightarrow P^{2-}(aq) + 2H_2O(l) \\ P^{2-}(aq) + OH^-(aq) &\leftrightarrow POH^{3-}(aq) \end{aligned} \qquad (3)$$

When the concentration of base is much larger than the concentration of anion $P^{2-}$, the second step in the reaction is pseudo first-order with respect to the anion, with a rate constant proportional to the concentration of base[35]. It may be shown that this step mimics the mineralization reaction for the carbon dioxide system in Figure 1, provided that it is accompanied by a small density decrease and the concentration of $POH^{3-}$ remains much smaller than the concentration of $P^{2-}$. Density measurements for this reaction in a well-mixed vessel connected to a capillary tube show that the former condition is satisfied, with $\rho_r \beta C'_s \approx 0.17 \, \mathrm{kg\,m^{-3}}$ for a hydroxide solution with 14<pH<14.7 and a concentration of phenolphthalein in MIBK of $5.5 \times 10^{-3}$ M. The flows were sufficiently slow for Taylor dispersion in the cell to be negligible[36].

Figure 4 shows the evolution of the pink diffusive layer of $P^{2-}$ as it grows into the underlying colourless hydroxide solution for three experiments with increasing values of $Da/Ra^2$. The convective fingers develop earlier and descend into the underlying solution faster for the slowly reacting case $a$ ($Da/Ra^2 = 5\times10^{-4}, t > 1.5\times10^2$) than for the intermediate reaction speed, case $b$ ($Da/Ra^2 = 1\times10^{-3}, t > 2.9\times10^2$), while convection occurs only during a short period of time for the fast reaction in case $c$ ($Da/Ra^2 = 2.7\times10^{-3}, 3.4\times10^2 < t < 1.5\times10^3$), after which the boundary layer remains diffusive and confined to the top of the hydroxide layer. Gradual fading of the pink fluid is observed in all three cases at longer times, showing the progression of the second step in reaction (3). These experimental observations indicate both a delay of onset and a weakening of convection as $Da/Ra^2$ increases, thus supporting our earlier theoretical findings. The measured times for onset of convection are compared non-dimensionally with the linear stability results in Figure 5, showing very good agreement. The very weak convection observed in case $c$ is well reflected by the drastic curbing effect of reaction on the predicted maximum growth rate $\sigma_{max\,kt}$ (over time and wavenumber). Also shown in Figure 5 is the onset time for the inert system consisting of acetic acid dissolved in MIBK overlying water. The wavenumber at the onset of instability is estimated for the reactive case $a$ to be $k \approx 0.03$, which is significantly smaller than the theoretical value of 0.06 (Figure 3), owing to the merger of initial instabilities before they become visible.

**Discussion.** A faster mineralization reaction between $CO_2$ dissolved in brine and the surrounding rock matrix enhances the trapping of $CO_2$ in solid form, thus increasing its long-term retention in the subsurface. However, as shown above, reaction may also curb significantly the downward convective transport of $CO_2$ in the reservoir, retaining it at shallower depths. For example, chemical reaction between the acidic brine and a rock formation rich in calcium feldspar (see reaction (2)), promotes the precipitation of calcite and kaolinite, removing carbon dioxide from the liquid phase and thereby reducing its density. Assuming the rock matrix has porosity of 0.1 and permeability of $10^{-13}$ m$^2$, we estimate $Da/Ra^2 = 2.6\times10^{-3}$ (using kinetic data from Marini[1]). Convection is predicted to initiate one month after injection of the carbon dioxide, with convections streams as wide as 1 m. However, these streams remain weak and completely shut-off after a short period of only two months. After this, the carbon dioxide will be transported by much slower diffusional processes. The results presented in this work have important practical implications for storage of carbon dioxide in saline aquifers, enabling informed screening of the most effective sites.

We should note that in most geological settings heterogeneity of the porous rock will have an important effect on buoyant fingering convection[10, 37-39]. Ennis-King and Paterson[38] and Rapaka *et al.*[39] have shown that anisotropy in sedimentary rocks leads to an increase of the time for onset of convection in a chemically-inert system by a factor of 10, for a vertical to horizontal permeability ratio of 0.1. Such an effect would reduce the period of convection and lead to an even earlier shut-off of vertical transport of carbon dioxide in a reactive system as considered here. Typical geological scales in layered systems range from metres to tens of metres (separation between flow barriers)[38]. These scales are larger than our estimate above of a diffusive lengthscale of 1 m, so we expect this flow to be mainly affected by the fine-scale permeability and anisotropy, but less so by a coarse layered structure[38]. The estimate of the period of convection for the basic homogeneous system presented here constitutes therefore an upper bound for more realistic heterogeneous rocks.

Our results also aid the understanding of the natural flow of carbon dioxide in the subsurface and its contribution to the global carbon cycle and, ultimately, climate change. The work is of broader relevance for diffusive boundary layers in porous media encountered in other diverse fields, including enhanced oil recovery[36], geothermal systems[40], and pollution and brine migration in the subsurface[41].

**Methods**

**Linear stability analysis.** The system in Figure 1 is described by the Darcy-Boussinesq[26,41] equations (4)-(6) for the fluid velocity $\mathbf{v}'\left[\text{LT}^{-1}\right]$, the reduced pressure $p' = P' - \rho_r g z' \left[\text{ML}^{-1}\text{T}^{-2}\right]$ and concentration of dissolved solute $C'\left[\text{ML}^{-3}\right]$:

$$\nabla' \cdot \mathbf{v}' = 0, \qquad (4)$$

$$\mathbf{v}' = -\frac{k}{\mu}\left(\nabla p' - \rho_r \beta C' g \mathbf{i}\right), \qquad (5)$$

$$\phi \frac{\partial C'}{\partial t'} + \mathbf{v}' \cdot \nabla' C' = D\phi \nabla'^2 C' - k_r a C'. \qquad (6)$$

We assume that diffusion dominates over convective dispersion for the slow flows considered in this study[22]. Here, the permeability of the porous medium $k\left[\text{L}^2\right]$ and the fluid viscosity $\mu\left[\text{ML}^{-1}\text{T}^{-1}\right]$ are constant. Density $\rho\left[\text{ML}^{-3}\right]$ is taken to depend linearly on $C'$ as $\rho = \rho_r(1+\beta C')$, where $\rho_r \left[\text{ML}^{-3}\right]$ is the reference density and $\beta \left[\text{M}^{-1}\text{L}^{3}\right]$ is the coefficient of density change due to concentration. The acceleration due to gravity is $g\left[\text{LT}^{-2}\right]$ and $\mathbf{i}$ is the unit vector co-directional with the positive vertical $z'$-axis; $z'$ increases downward and has origin at the interface between the supercritical carbon-dioxide and the brine layers. Time is denoted by $t'[\text{T}]$. We note that when dissolution of the solute increases the density of the fluid, $\beta$ is positive; reaction counteracts the dissolution effect on density by consuming solute, thereby having a stabilizing role on the diffusive layer for the configuration in Figure 1. Introducing the scales $L_\text{C} = \mu D\phi/(k\Delta\rho_0 g)$, $v_\text{s} = D\phi/L_\text{C}$, $t_\text{s} = L_\text{C}^2/D$, $p_\text{s} = \mu D\phi/k$, and the solubility of the solute in the fluid $C'_s$, the governing equations are nondimensionalized, giving

$$\nabla \cdot \mathbf{v} = 0, \qquad (7)$$

$$\mathbf{v} = -\nabla p + C\,\mathbf{i}, \qquad (8)$$

and
$$\frac{\partial C}{\partial t} + \mathbf{v}\cdot\nabla C = \nabla^2 C - \frac{\text{Da}}{\text{Ra}^2}C. \qquad (9)$$

Thus, the only parameter determining flow and transport in this system is a ratio of the Damköhler number and the square of the solutal Rayleigh-Darcy number, $\text{Da}/\text{Ra}^2 = k_r a D\phi/(k\Delta\rho_0 g/\mu)^2$. Here, $\Delta\rho_0 = \rho_r \beta C'_s$ is the maximum density contrast between pure and solute-saturated fluid. The system of equations (7)-(9), with initial condition $C_b(0,z) = 0$, far field condition $C_b(t, z\to\infty) = 0$ and interface condition $C_b(t,0) = 1$, has a stagnant base state

$$C_b(t,z) = \exp\left(-\frac{\text{Da}}{\text{Ra}^2}t\right)\text{erfc}\left(\frac{z}{2\sqrt{t}}\right) + \frac{\text{Da}}{\text{Ra}^2}\int_0^t \exp\left(-\frac{\text{Da}}{\text{Ra}^2}s\right)\text{erfc}\left(\frac{z}{2\sqrt{s}}\right)ds \qquad (10)$$

$$\mathbf{v}_b = 0, \quad p_b = p_b(z,t), \qquad (11)$$

where $p_b(z,t)$ is the solution of equation (8) for the concentration profile in (10). Linearizing equations (7)-(9) around the base state (10)-(11), eliminating pressure and horizontal velocity, and expanding transverse perturbations into time-dependent Fourier modes as $\{\hat{v}(x,\eta,t), \hat{C}(x,\eta,t)\} = \{v_0(\eta,t), C_0(\eta,t)\}e^{ikx}$ leads to,

$$\frac{1}{4t}\frac{\partial^2 v_0}{\partial \eta^2} - k^2 v_0 = -k^2 C_0 \qquad (12)$$

$$t\frac{\partial C_0}{\partial t} = \left(\frac{1}{4}\frac{\partial^2}{\partial \eta^2} + \frac{\eta}{2}\frac{\partial}{\partial \eta}\right)C_0 - \left(k^2 + \frac{\text{Da}}{\text{Ra}^2}\right)C_0 t - \frac{v_0}{2}\sqrt{t}\frac{\partial C_b}{\partial \eta} \qquad (13)$$

where $\eta = z/(2\sqrt{t})$ and $k$ is the wavenumber in the horizontal $x$-direction. The boundary conditions are $v_0 = C_0 = 0$ at $\eta = 0, +\infty$. This initial-value problem was solved numerically using the finite-element method in partial-differential-equation solver Fastflo[42]. Test simulations were conducted to ensure that the results became independent of particular initial conditions well below the times for onset of instability and agreed with previously published solutions for the inert system[20]. The concentration perturbation was measured using the norm $\overline{C}_0(t) = \int_0^{L_\eta} C_0(\eta,t)d\eta/L_\eta$, where $L_\eta$ is the extent of the computational domain. The growth rate of this perturbation was defined as $\sigma = \ln\left[\overline{C}_0(t)/\overline{C}_0(t-\Delta t)\right]/\Delta t$.

**Experimental setup.** The experiments were performed in a vertically oriented Hele-Shaw cell, split into upper and lower hemi-cells (width 120 mm × height 95 mm × gap thickness 1 mm) based on the design of Kuster et al.[34]. The upper cell slid down two stainless steel rails to rest on the lower cell, allowing two reacting fluids to come into contact with minimum disturbance to the interface. A solution of phenolphthalein dissolved in 4-methyl-2-pentanone (MIBK), with a concentration of $5.5 \times 10^{-3}$ M, was injected into the upper cell; this solution was held in the cell by surface tension. An aqueous solution of sodium hydroxide was placed in the lower cell. Experiments were performed for sodium hydroxide concentrations of 1.0 M, 2.0 M and 5.0 M, corresponding to $\text{Da}/\text{Ra}^2$ of approximately $5.0 \times 10^{-4}$, $1.0 \times 10^{-3}$ and $2.7 \times 10^{-3}$. For the inert system, with $\text{Da}/\text{Ra}^2 = 0$, the upper layer consisted of acetic acid dissolved in MIBK with concentrations in the range 0.10 - 1.7 M, while the lower layer was water.

Chemical compatibility between the cell plates and the fluids was ensured by using upper and lower plates of, respectively, borosilicate glass and metacrylate plastic for the reactive system; for the inert system, all plates were of borosilicate glass. Uniform backlighting from diffuser light boxes allowed camera recording of front-view images of the onset and development of convection in the lower layer, while the upper layer remained stagnant. The times for onset of convection for the inert and reactive systems were determined from image analysis.

**Acknowledgements**

J.T.H.A. gratefully acknowledges the Schlumberger Foundation for financial support for her PhD study. S.S.S.C. and J.T.H.A. thank PhD student P. Ghoshal for the measurement of density change in the reactive experiments.

**Author contributions**

S.S.S.C. and J.T.H.A. designed the study; J.T.H.A. performed the experiments and data analysis; S.S.S.C. performed the theoretical modelling and prepared the manuscript.

**Additional information**

Competing financial interests: The authors declare no competing financial interests.

(FIGURES AND FIGURE LEGENDS)

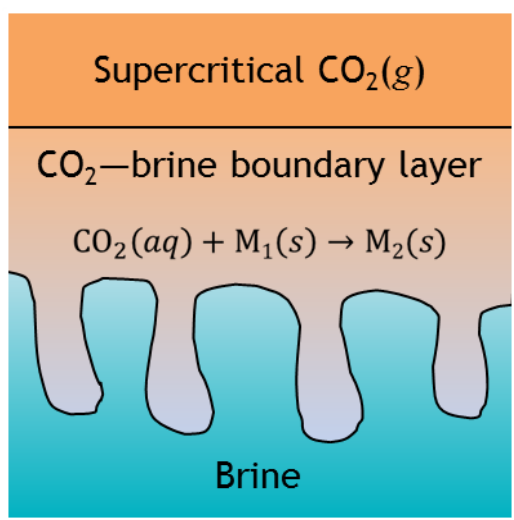

FIG. 1. **Onset of buoyant fingering.** Carbon dioxide dissolves and diffuses into the underlying brine in a porous saline aquifer. The layer of dense $CO_2$-rich brine eventually breaks up into fingers that grow and descend further into the brine. Geochemical reactions between the acidic $CO_2$-rich brine and aquifer rock may occur. E.g., $M_1$ might represent calcium feldspar and $M_2$ is then a mixture of calcite and kaolinite.

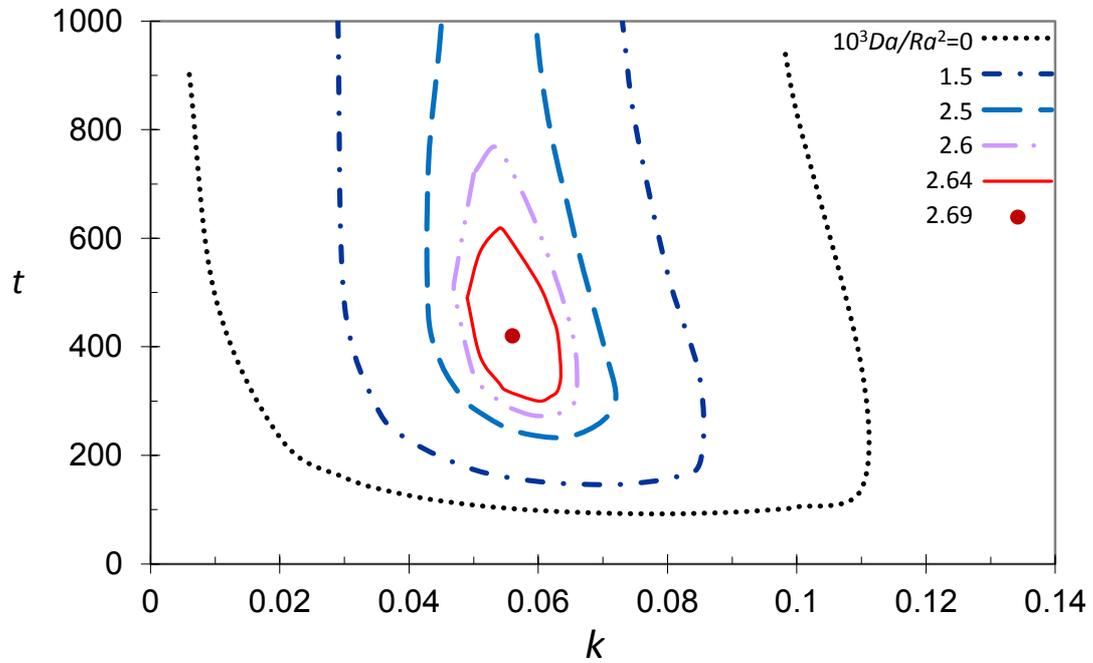

FIG. 2. **Speed of geochemical reaction affects the onset of convection.** As the reaction speed ($Da/Ra^2$) increases, the time for onset of convection $t$ increases and the range of unstable transverse wavenumbers $k$ decreases ($t$ and $k$ are non-dimensional). For sufficiently fast reactions, convection shuts off at large times.

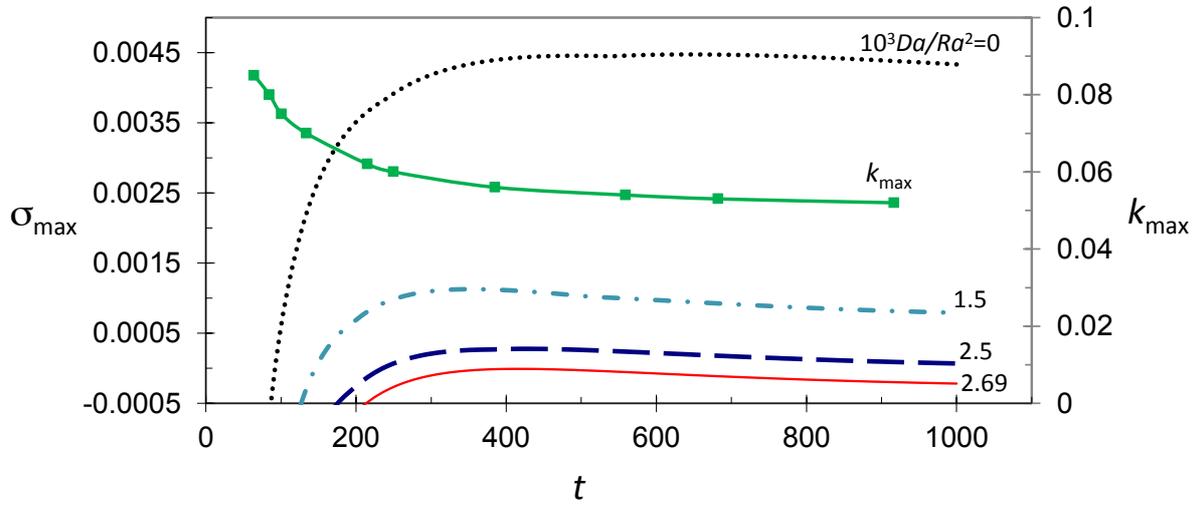

FIG. 3. **Speed of geochemical reaction affects the growth rate and pattern of the instability.** Non-dimensional evolution of the maximum growth rate $\sigma_{max}$ and most unstable wavenumber $k_{max}$ for different reaction speeds. A rise in reaction speed causes a pronounced decrease in the maximum growth rate of the fingers and a decrease in the wavenumber at onset of convection.

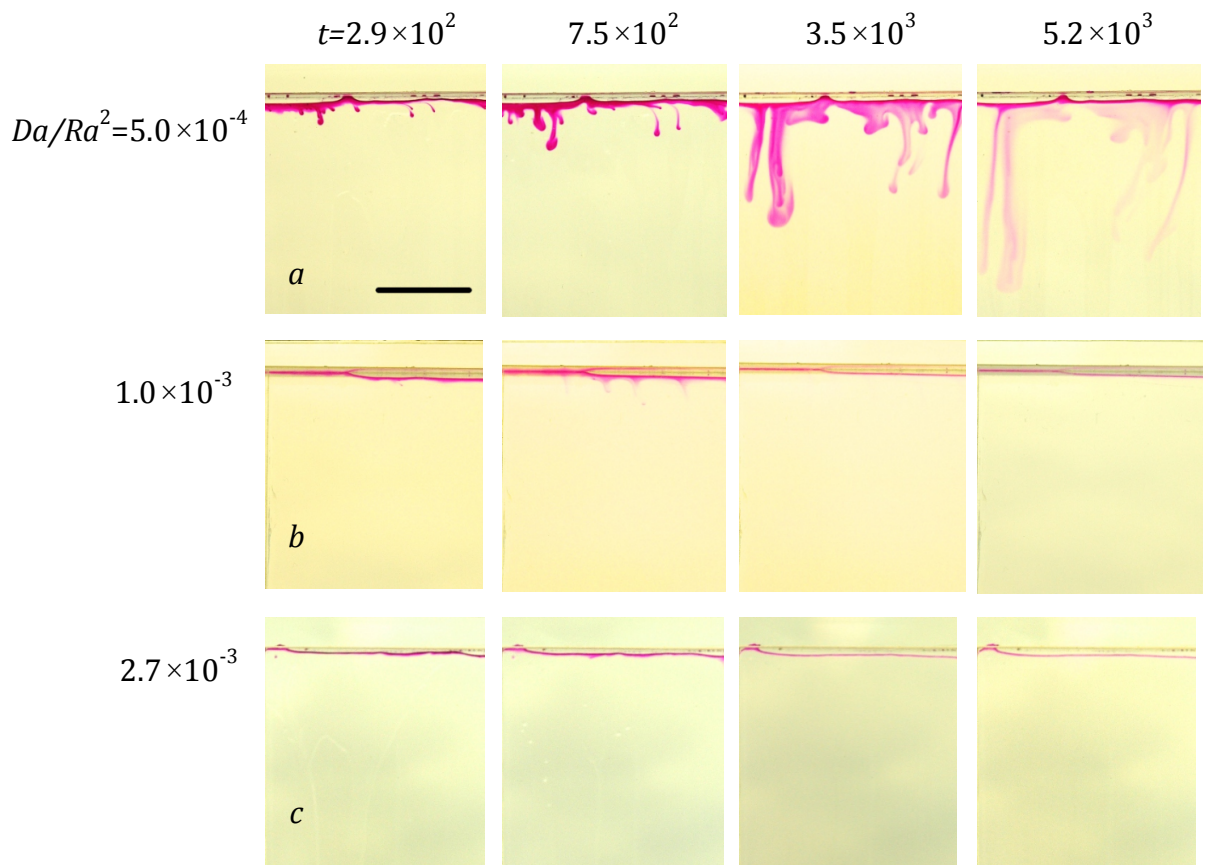

FIG. 4. **Laboratory experiments demonstrating the shut off of convection by chemical reaction.** Sequence of photographs showing the evolution of the pink diffusive boundary layer formed by the reaction of phenolphthalein in MIBK (top layer) with aqueous sodium hydroxide (bottom layer) in a Hele-Shaw cell; the focus is on the lower aqueous layer. Reaction speed ($Da/Ra^2$) increases from cases *a* to *c*. Convective fingering is more vigorous for the slowly reacting case *a* than for the intermediate reaction speed in case *b*, while motion is very weak in the presence of the fast reaction in case *c*. Time *t* is non-dimensional. Scale bar, 2 cm. The background colour of each picture was lightened using *iPhoto*.

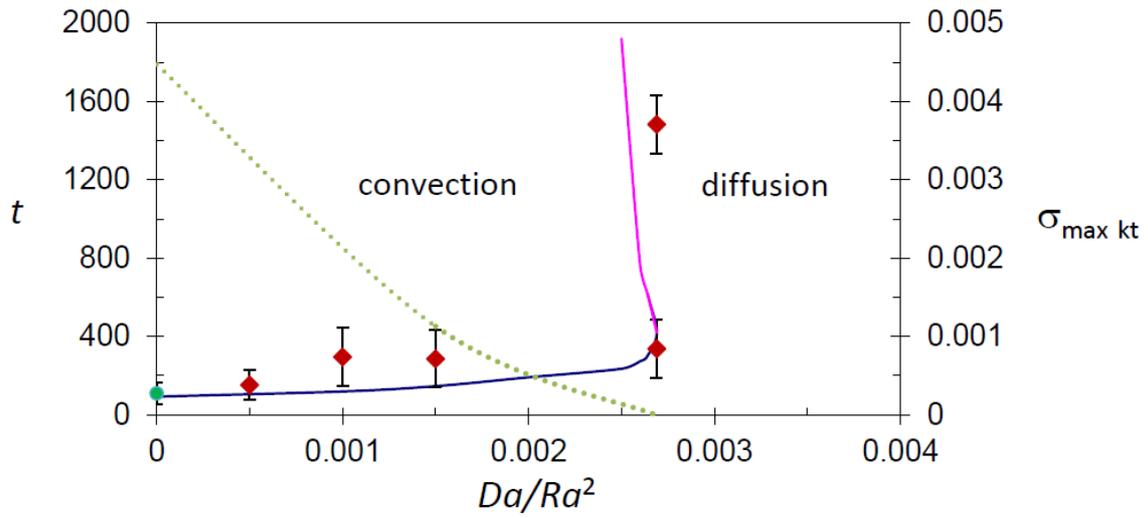

FIG. 5. **Regimes of transport of dissolved solute.** Effects of reaction speed $Da/Ra^2$ on times for onset (blue solid line) and cessation of convection (pink solid line), as well as on the maximum growth rate $\sigma_{max\,kt}$ (dotted line) predicted from linear stability analysis.

Experimental measurements of times for onset and cessation of convection for the reactive (diamonds) and inert (circle) systems are also shown. Time $t$ is non-dimensional. The error bars represent the standard deviation of three experimental measurements. The solid lines separate the conditions for dominant transport of solute by convective motion and for transport by molecular diffusion.